\documentstyle[epsf]{article}
\input epsf
\begin{document}
\title{{\bf The role of g-wave pairing and Josephson tunneling
in high-$T_c$ cuprate superconductors.}}
\author{\Large P.V.Shevchenko$^{a}$, O.P.Sushkov$^{b}$\\
School of Physics, The University of New South Wales\\
Sydney 2052, Australia}
\maketitle
\begin{abstract}
The implications of the two-pocket Fermi surface for macroscopic quantum
phenomena are considered. We demonstrate that in the case of the two-pocket 
Fermi surface the g-wave pairing is closely related to the d-wave one. As a 
result two macroscopic condensates arise. The Josephson tunneling for such 
two-component system has very special properties. We prove that the presence 
of the g-wave does not contradict the existing experimental data 
on tunneling. We also discuss the possible ways to experimentally reveal 
the g-wave component.
\end{abstract}
\large
\section{Introduction}
There is a lot of controversy about the shape of the Fermi surface in
cuprate superconductors.  In the early days it was believed that it is a
small Fermi surface of the doped Mott insulator \cite{Bed}. Later
many of the results from photoelectron spectroscopy (PES) have been
interpreted as favoring a large Fermi surface in agreement with
Lattinger's theorem \cite{Dessau}. On the other hand most recent PES
data\cite{Aebi,Chak,Wells,Mars} once more give indications of a small Fermi
surface for underdoped samples.
In the present paper we consider the scenario with a small
Fermi surface consisting of hole pockets around $(\pm \pi/2, \pm \pi/2)$,
see Fig.1a.
It is widely believed that the $t-J$ model
describes the main details of the doped Mott insulator. To fit the experimental
hole dispersion one needs to extend the model introducing additional hopping 
matrix elements $t^{\prime}$, $t^{\prime \prime}$ 
(see. e.g. Refs.\cite{Naz,Bala,Sus}), but basically it is the $t-J$ model. 
Superconducting pairing in the $t-J$ model induced
by spin-wave exchange has been considered in the papers \cite{Flam,Bel}.
It was demonstrated \cite{Flam}  that there is an infinite set of
solutions for the superconducting gap. All the solutions have nodes
along the lines $(1,\pm 1)$, see Fig.1a. (It is very convenient to
use the magnetic Brillouin zone, but it can certainly  be mapped
to the full zone.) Using translation by the vector of inverse magnetic lattice
the picture can be reduced to two hole pockets centered around
the points $(\pm \pi/2, \pi/2)$, see Fig.1a. The superconducting pairing is
the strongest between particles from the same pocket, and the lowest energy
solution for the superconducting gap has only one node line in each pocket.
Having this solution in a single pocket one can generate two solutions in
the whole Brillouin zone taking symmetric or antisymmetric combinations
between pockets. The symmetric combination corresponds to the d-wave
(Fig.1b), and the antisymmetric combination corresponds to the g-wave (Fig.1c)
pairing. 
We would like to note that the possibility to generate new solutions taking
different combinations between pockets was first demonstrated
by Scalettar, Singh, and Zhang in the paper \cite{Scal}.
The energy splitting between the d- and g-wave solutions has been
investigated numerically in the Ref.\cite{Bel}.
The g-wave solution disappears and only d-wave one survives
as soon as the hole dispersion is degenerate along the face of the magnetic
Brillouin zone. Actually in this situation there are no pockets
and one has a large open Fermi surface at an arbitrary small hole 
concentration.
However for small well separated pockets the d- and g-wave solutions are almost
degenerate. In pure $t-J$ or $t-t^{\prime}-t^{\prime \prime}-J$ model the
d-wave solution always has the lower energy. However if one extends the model
including nearest sites hole-hole Coulomb repulsion the situation can be
inverted. The nearest sites repulsion does not influence the  g-wave pairing 
and substantially suppresses the d-wave pairing. So it is quite possible that 
the real ground state has the g-wave superconducting gap.

\section{Ginzburg-Landau free energy}

In the present paper we consider the scenario with hole pockets
well separated in k-space. According to the described above microscopic 
picture we should consider simultaneously the d- and the g-wave pairing. In 
general terms such situation was considered a long time ago for conventional 
superconductors by Leggett \cite{Leg}. Let us formulate an effective
Ginzburg-Landau (GL) theory.
In the first approximation we can neglect the interaction between 
pockets and introduce two macroscopic condensates corresponding to the
pockets,
$\Psi_1=\left|{\Psi_1}\right|\mbox {e}^{i\phi_1}$,
$\Psi_2=\left|{\Psi_2}\right|\mbox{e}^{i\phi_2}$,
where $\left|{\Psi_1}\right|=\left|{\Psi_2}\right|=\left|\Psi\right|=
\sqrt{N_h}/2$,
$N_h$ is the number density of the condensate holes.
The Ginzburg-Landau free energy of a such system in an external magnetic
field can be written as
\begin{eqnarray}
\label{F0}
F_0&=&\int\left(
{1\over{2m^{\star}}}|(\nabla-\frac{2ie}{\hbar c}{\bf A})\Psi_1|^2+
{1\over{2m^{\star}}}|(\nabla-\frac{2ie}{\hbar c}{\bf A})\Psi_2|^2-\right.\\
&&\left. a{\left|\Psi_1\right|}^2+b{\left|\Psi_1\right|}^4-
a{\left|\Psi_2\right|}^2+b{\left|\Psi_2\right|}^4+\frac{B^2}{8\pi}\right)dV
\nonumber.
\end{eqnarray}
A small interaction between pockets can be described by adding to the free
energy (\ref{F0}) the term
\begin{equation}
\label{Fint}
F_{\mbox{{\small int}}}=
\gamma \int \left(\Psi_1^*\Psi_2+\Psi_1\Psi_2^*\right)dV \to
2\gamma V \left|\Psi_1\right|\left|\Psi_2\right|\cos(\phi_1-\phi_2),
\end{equation}
where $\gamma$ is a small parameter of the interaction, $\gamma \ll a$.
The total bulk energy of the cuprate superconductor equals
$F_{V}^{dg}=F_0+F_{\mbox{{\small int}}}$, and
the equilibrium values of the order parameters are
\begin{equation}
\label{m}
\left|\Psi_1\right|^2=\left|\Psi_2\right|^2=\left|\Psi\right|^2=
(a+\left|\gamma\right|)/2b\approx a/2b.
\end{equation}
The ground state phase difference $\Delta \phi= \phi_2 -\phi_1$
is determined by the sign of $\gamma$:
if $\gamma>0$, then $\Delta \phi=\pi$ ( g-wave );
if $\gamma<0$, then $\Delta \phi=0$ ( d-wave ).
It is also convenient to introduce the d- and g-wave condensates
$\Psi_d=\Psi_1+\Psi_2=2|\Psi|\mbox{cos}(\Delta\phi/2)e^{i\phi}$ and
$\Psi_g=\Psi_1-\Psi_2=2|\Psi|\mbox{sin}(\Delta\phi/2)e^{i\phi-i\pi/2}$,
where $2\phi=\phi_1+\phi_2$. So the ground state has either d- or g-wave 
symmetry: if $\gamma>0$ then $\Psi_d=0, \Psi_g\neq 0$, if 
$\gamma<0$ then $\Psi_g=0, \Psi_d\neq 0$.

\section{Tunnel junction}

Let us consider a Josephson tunnel contact of a conventional s-wave
superconductor (the order parameter 
$\Psi_s=\left|\Psi_s\right|\mbox{exp}{(i\phi_s)}$) and a high-$T_c$ cuprate
superconductor.
Constant supercurrent ${\bf j}$ is maintained through the contact.
The bulk free energy of the conventional superconductor equals
\begin{equation}
\label{Fsv}
F_V^{s}=\int\left(\frac {\hbar^2}{2m_s^{\star}}\left|({\bf \nabla}  - \frac{2ie}{\hbar c}{\bf A})\Psi_s
\right|^2-a_s\left|\Psi_s\right|^2+b_s\left|\Psi_s\right|^4+\frac{B^2}{8\pi}\right)dV,
\end{equation}
and the surface free energy related to the contact is
\begin{equation}
\label{Fc}
F_S=\int \left(-\lambda_1({\bf n})\left[\Psi_{s-}\Psi_{1+}^*+\Psi_{s-}^*
\Psi_{1+}\right]-\lambda_2({\bf n})\left[\Psi_{s-}\Psi_{2+}^*+\Psi_{s-}^*\Psi_
{2+}\right] \right)dS,
\end{equation}
where {\bf n} is a unit vector orthogonal to the surface
of the contact and directed from conventional superconductor to cuprate,
$\lambda_1({\bf n})$ and $\lambda_2({\bf n})$
are the tunneling matrix elements from the first and the second pockets correspondingly.
$\xi$ is an axis parallel to ${\bf n}$, and $\Psi_{1+}$, $\Psi_{2+}$, and $\Psi_{s-}$ 
are values of the condensates at the contact ($\xi=\pm 0$). 
It is convenient to use following simple parameterization for the tunneling
matrix elements
\begin{eqnarray}
\label{lam}
\lambda_1({\bf n})&=&C(n_x^2-n_y^2)e^{-\alpha(n_x-n_y)^2},\\
\lambda_2({\bf n})&=&C(n_x^2-n_y^2)e^{-\alpha(n_x+n_y)^2},\nonumber
\end{eqnarray}
where $C$ and $\alpha$ are some parameters depending on the tunneling probability,
the shape of the Fermi surface, etc. The axes x and y are directed along crystal 
axes a and b of the cuprate. We stress that eq.(\ref{lam}) is just a 
parameterization having correct symmetry properties. Let us introduce also the d- 
and g-wave tunneling matrix elements: 
$\lambda_d({\bf n})=\lambda_1({\bf n})+\lambda_2({\bf n})$ and 
$\lambda_g({\bf n})=\lambda_1({\bf n})-\lambda_2({\bf n})$.
The symmetries of these matrix elements shown at Fig.2 are the same as the
symmetries of d- and g-wave gaps in the momentum space, Fig. 1b,c.
After variation of the total free energy $F=F_{V}^{dg}+F_V^s+F_S$
with respect to $\Psi_1^*,\Psi_2^*,\Psi_s^*$ and vector potential ${\bf A}$
we find:\\
the GL equations for conventional and high-$T_c$ superconductors
\begin{eqnarray}
\label{GL}
&&-\frac{\hbar^2}{2m_s^{\star}}\left({\bf \nabla} -\frac{2ie}{\hbar c}{\bf A}
\right)^2\Psi_s-a_s\Psi_s+2b_s\left|\Psi_s\right|^2\Psi_s=0,\nonumber\\
&&-\frac{\hbar^2}{2m^{\star}}\left({\bf \nabla} -\frac{2ie}{\hbar c}{\bf A}
\right)^2\Psi_1-a\Psi_1+2b\left|\Psi_1\right|^2\Psi_1+\gamma \Psi_2=0,\\
&&-\frac{\hbar^2}{2m^{\star}}\left({\bf \nabla} -\frac{2ie}{\hbar c}{\bf A}
\right)^2\Psi_2-a\Psi_2+2b\left|\Psi_2\right|^2\Psi_2+\gamma \Psi_1=0,\nonumber
\end{eqnarray}
the boundary conditions for these equations
\begin{eqnarray}
\label{b}
-\left. \frac{\hbar^2}{2m_s^{\star}}{\bf n}\left({\bf \nabla} -\frac{2ie}{\hbar c}
{\bf A}\right)\Psi_s\right|_{\xi=-0}&=&
\lambda_1({\bf n})\Psi_{1+}+\lambda_2({\bf n})\Psi_{2+},\nonumber\\
\left. \frac{\hbar^2}{2m^{\star}}{\bf n}\left({\bf \nabla} -\frac{2ie}{\hbar c}
{\bf A}\right)\Psi_1\right|_{\xi=+0}&=&\lambda_1({\bf n})\Psi_{s-},\\
\left. \frac{\hbar^2}{2m^{\star}}{\bf n}\left({\bf \nabla} -\frac{2ie}{\hbar c}
{\bf A}\right)\Psi_2\right|_{\xi=+0}&=&\lambda_2({\bf n})\Psi_{s-},\nonumber
\end{eqnarray}
and the currents in the cuprate and conventional  superconductors
\begin{eqnarray}
\label{j}
{\bf j}&=&-\frac{\hbar i e}{m^{\star}}\left(\Psi_1^*{\bf \nabla} \Psi_1-\Psi_1
{\bf \nabla} \Psi_1^*+
\Psi_2^*{\bf \nabla} \Psi_2-\Psi_2\nabla \Psi_2^*\right)-\frac{4e^2}{cm^{\star}}
{\bf A}(\left|\Psi_1\right|^2+\left|\Psi_2\right|^2),\\
{\bf j}&=&-\frac{\hbar i e}{m_s^{\star}}\left(\Psi_s^*{\bf \nabla} \Psi_s-\Psi_s 
{\bf \nabla} \Psi_s^*
\right)-\frac{4e^2}{cm_s^{\star}}{\bf A}\left|
\Psi_s\right|^2.\nonumber
\end{eqnarray}
Substituting the boundary conditions (\ref{b}) into eqs. (\ref{j})
we find the expression for supercurrent through the contact
\begin{equation}
\label{jc}
j=\frac{2ie}{\hbar}\lambda_1({\bf n})[\Psi_{s-}\Psi_{1+}^*-\Psi_{s-}^*
\Psi_{1+}]+\frac{2ie}{\hbar}\lambda_2({\bf n})[\Psi_{s-}\Psi_{2+}^*-
\Psi_{s-}^*\Psi_{2+}].
\end{equation}

We will consider the distances larger than the cuprate superconducting correlation 
length. At these distances the magnitudes $\left|\Psi_1\right|$ and
$\left|\Psi_2\right|$ are equal to the constant given by eq.(\ref{m}), and
only the phases of condensates are $\xi$ dependent.
The Josephson current through the contact can be written as
\begin{equation}
\label{jf}
j=\frac{4e}{\hbar}\left|\Psi\right|\left|\Psi_s\right|[\lambda_1({\bf n})
\sin(\phi_{1+}-
\phi_{s-})+\lambda_2({\bf n})\sin(\phi_{2+}-\phi_{s-})],
\end{equation}
where $\phi_{s-}$, $\phi_{1+}$ and $\phi_{2+}$ are values of the phases
at the contact ($\xi= \pm 0$).
We consider the situation without an external magnetic field in the contact.
The magnetic field caused by the supercurrent is very small, therefore we
set ${\bf A}=0$ and reduce eqs.(\ref{GL}) for $\Psi_1$ and
$\Psi_2$ to ones for the phases
\begin{eqnarray}
\label{GL1}
&&\frac{\hbar^2}{2m^{\star}}\frac{d^2\Delta\phi}{d\xi^2}+
i\frac{\hbar^2}{m^{\star}}\frac{d\phi}
{d\xi}\frac{d\Delta\phi}{d\xi}+2\gamma\sin\Delta\phi=0,\\
&&\frac{\hbar^2}{m^{\star}}\left[\left(\frac{d\phi}{d\xi}\right)^2+\frac{1}{4}
\left(\frac{d\Delta\phi}{d\xi}\right)^2\right]-i\frac{\hbar^2}{m^{\star}}\frac{d^2
\phi}{d\xi^2}+2\left|\gamma\right|(1+\frac{\gamma}{\left|\gamma\right|}
\cos\Delta\phi)=0,\nonumber
\end{eqnarray}
with $\phi=(\phi_1+\phi_2)/2$ and $\Delta\phi=\phi_2-\phi_1$.
The equilibrium value of $\Delta\phi$ is $\Delta\phi_0$ with 
$\Delta\phi_0=0$ at $\gamma < 0$ ( d-wave ), and
$\Delta\phi_0=\pi$ at $\gamma > 0$ ( g-wave ).
If the current is small we can linearize eqs.(\ref{GL1}) with respect to
$\Delta\phi-\Delta\phi_0$. This gives that $\phi=(\phi_1+\phi_2)/2=
\mbox{const}$, and 
\begin{equation}
\label{D}
\Delta \phi=\phi_2-\phi_1=\Delta\phi_0+A\exp\left(-\xi/l_{\gamma}\right),
\end{equation}
where $l_{\gamma}=\hbar/\sqrt{4\left|\gamma\right|m^{\star}}$.
It is convenient to rewrite the supercurrent (\ref{jf}) in the form
\begin{equation}
\label{jff}
j=\frac{4e}{\hbar}\left|\Psi\right|\left|\Psi_s\right|\left[
\lambda_d \sin\theta \cos\frac
{\Delta\phi_c}2-\lambda_g\cos\theta\sin\frac{\Delta\phi_c}{2}\right],
\end{equation}
where $\theta =\phi-\phi_{s-}$ and $\Delta\phi_c$ is the value of $\Delta\phi$
at the contact: $\Delta\phi_c=A$ for $\gamma < 0$ ( d-wave ) and
$\Delta\phi_c=\pi+A$ for $\gamma > 0$ ( g-wave ).
The solution we found depends on the two arbitrary constants $\theta$ and $A$.
If the supercurrent through the contact is fixed then the eq. (\ref{jff}) 
gives one relation between these constants.
Second relation follows from the minimum of  total free energy.
After substitution of the solution (\ref{D}) into expressions (\ref{F0}),(\ref{Fint}),
(\ref{Fc}) we find the part of the total free energy depending on $\theta$ and $A$
\begin{equation}
\label{FO}
F_{\Omega}=\left|\Psi\right|^2K_{\gamma}A^2-
2\left|\Psi\right|\left|\Psi_s\right|\left[ \lambda_d\cos\theta \cos
\frac{\Delta\phi_c}2+\lambda_g\sin\theta \sin\frac{\Delta\phi_c}2\right].
\end{equation}
Here $K_{\gamma}=l_{\gamma}\left|\gamma\right|=
\hbar\sqrt{\left|\gamma\right|/m^*}/2$.

Let us consider at first the contact plane to be parallel to the axis a or axis b of
the cuprate (${\bf n}=(\pm 1,0)$ or ${\bf n}=(0,\pm 1)$). In this case $\lambda_g=0$ and 
$\lambda_d \ne 0$. If the ground state
pairing has d-wave symmetry ($\gamma < 0$) we find from eqs. (\ref{jff}) and 
(\ref{FO}) the following condition of minimum
\begin{equation}
\label{min1}
\left. \frac{d F_{\Omega}}{d A}\right|_{j=const}=\left(
2K_{\gamma}\left|\Psi\right|^2+\frac 
{\lambda_d\left|\Psi_s \Psi\right|}{2\cos\theta}\right)A=0.
\end{equation}
Therefore $A=0$, and $\theta$ is determined by the usual relation
\begin{equation}
\label{ju}
j=j_{cd} \sin\theta, \ \ \ j_{cd}=\left|\frac{4e}{\hbar}\Psi \Psi_s\lambda_d\right|.
\end{equation}
If the ground state pairing has g-wave symmetry ($\gamma > 0$) the condition of 
minimum is
\begin{equation}
\label{min2}
\left. \frac{d F_{\Omega}}{d A}\right|_{j=const}=
\frac{\lambda_d\left|\Psi_s\right|}{\mbox{cos}\theta}+2A
K_{\gamma}\left|\Psi\right|=0.
\end{equation}
This gives 
\begin{equation}
\label{A}
A=-{{\lambda_d\left|\Psi_s\right|}\over{2\cos\theta K_{\gamma}\left|\Psi\right|}},
\end{equation}
and the supercurrent is
\begin{equation}
\label{jtan}
j={{e \lambda_d^2 \left|\Psi_s\right|^2}\over{\hbar K_{\gamma}}}\tan\theta.
\end{equation}
One has to remember that the above equations are derived assuming that $A \ll 1$ and
therefore they are not valid when $\theta$ is approaching $\pi/2$.
So the picture is that deep inside the cuprate we have g-wave pairing, but in the
layer of width $l_{\gamma}$ near the contact there is admixture of the d-wave
component. If current is increasing the admixture is increasing too.
So the current pumps the g-wave into the d-wave in the layer of width $l_{\gamma}$.
At critical current there is only the d-wave at the contact. Therefore the critical
current in this case is exactly the same as one for the d-wave ground state:
$j_c=j_{cd}$.

At an arbitrary orientation of contact plane with respect to the crystal axes
both  $\lambda_d$ and $\lambda_g$ do not vanish. Nevertheless the d-wave tunneling
probability is always larger than the g-wave one:
$\left|\lambda_d({\bf n})\right|>\left|\lambda_g({\bf n})\right|$. It is 
obvious from eqs.(\ref{lam}).
Let us rewrite the current (\ref{jff}) as
\begin{equation}
\label{j12}
j=\frac{2\sqrt{2}e}{\hbar}\left|\Psi\right|\left|\Psi_s\right|
\sqrt{\lambda_d^2+\lambda_g^2+(\lambda_d^2-\lambda_g^2)\cos\Delta\phi_c}
\cdot \sin(\theta-\alpha),
\end{equation}
where $\sin\alpha=\sqrt{2}\lambda_g\sin\frac{\Delta\phi_c}{2}\left/
\sqrt{\lambda_d^2+\lambda_g^2+(\lambda_d^2-\lambda_g^2)\cos\Delta\phi_c}
\right.$, and let us introduce the notations
$j_{cd}=\left|\frac{4e}{\hbar}\Psi \Psi_s \lambda_d\right|$, 
$j_{cg}=\left|\frac{4e}{\hbar}\Psi \Psi_s \lambda_g\right|$,
keeping in mind that $j_{cd} > j_{cg}$.
Let us assume first that the ground state of the cuprate has the d-wave gap 
($\gamma < 0$).
Then at any supercurrent the system is in the d-wave everywhere including the contact
($\Delta \phi_c = 0$), and from eq.(\ref{j12}) we conclude that the critical 
current is $j_c=j_{cd}$. 
Consider now the situation when the ground state of the cuprate has the g-wave 
symmetry ($\gamma > 0$). Then at $j < j_{cg}$ the system remains in g-wave 
state everywhere including the contact, $\Delta \phi_c = \pi$.
Increasing the current above $j_{cg}$ we start to pump the g-wave into the d-wave 
in the surface layer of the width $\l_{\gamma}$. When the current reaches
$j_{cd}$, according to eq.(\ref{j12}) the phase takes the value 
$\Delta \phi_c = 0$,
and this is the critical point. So at the critical point in the contact the g-wave is
completely pumped into the d-wave, and the critical current is exactly the same as 
in the case of the d-wave pairing: $j_c=j_{cd}$.

\section{SQUID}
The g-wave pairing can be revealed in the SQUID.
Consider at first the geometry of the SQUID experiment \cite{Harl}, results of 
which are interpreted as a very
strong evidence in favor of the d-wave pairing. The setup is shown schematically at
Fig.3. Faces of the superconducting corner are parallel  to crystal axes {\bf a}
and {\bf b} of the
cuprate superconductor so that $\lambda_g({\bf n})=0$  and $\lambda_d({\bf n})$ have 
opposite signs at the tunnel contacts:
$\lambda_{d1}=-\lambda_{d2}=\lambda_d$. The indexes 1,2 numerate the contacts.
Using eqs.(\ref{jff}),(\ref{FO}) the total current and the free energy can be written as
\begin{equation}
\label{jSQ}
j=j_1+j_2=\frac{4e}{\hbar}\left|\Psi\right|\left|\Psi_s\right|\lambda_d\left[
\sin\theta_1\cos\frac{\Delta\phi_{c1}}2-\sin\theta_2\cos\frac{\Delta\phi_{c2}}2\right],
\end{equation}
\begin{equation}
\label{FSQ}
F_{\Omega }=
\left|\Psi\right|^2 K_{\gamma}(A_1^2+A_2^2)-
2\left|\Psi\right|\left|\Psi_s\right|\lambda_d\left[
\cos\theta_1\cos\frac{\Delta\phi_{c1}}2-\cos\theta_2
\cos\frac{\Delta\phi_{c2}}2\right].\nonumber
\end{equation}
Here $\theta_i$, $A_i$, and $\Delta\phi_{ci}=\Delta \phi_0+A_i$ (i=1,2) 
are the quantities introduced in previous
section. Quantization condition is standard
\begin{equation}
\label{phiq}
\theta_1=\theta+{{\pi}\over{2}}-\pi\frac{\Phi}{\Phi_0},\hspace{0.2cm} 
\theta_2=\theta+{{\pi}\over{2}}+\pi\frac{\Phi}{\Phi_0}-2\pi m,
\end{equation}
where $\Phi_0=\pi c \hbar/e$ is the quantum of magnetic flux, and $m$ is an integer 
number. 

Let us consider at first the case $\gamma<0$ which corresponds to the
d-wave pairing in the cuprate ground state.
Minimum of  the free energy (\ref{FSQ}) at fixed supercurrent (\ref{jSQ})
is defined by the equations 
\begin{equation}
\label{A12}
\left.\frac{\partial F_{\Omega}}{\partial A_i}\right|_{j=const}=
\left(2K_{\gamma}\left|\Psi\right|^2+{{\lambda_d \left|\Psi \Psi_s\right|
\sin \left(\pi \Phi/\Phi_0 \right)}\over{2 \cos \theta}}\right)A_i=0.
\end{equation}
The solution of these equations is $A_1=A_2=0$, so the system remains in the d-wave
state everywhere including the contacts, and the current equals
\begin{equation}
\label{jcs}
j=j_{cd} \sin\theta, \ \ \ j_{cd}=\frac{8e}{\hbar}\left|\Psi \Psi_s \lambda_d\right|
\left|\sin\left(\pi\frac{\Phi}{\Phi_0}\right)\right|.
\end{equation}
The plot of $j_{cd}$ is represented at Fig.4a and is in agreement with experimental
data [13].\\
Now consider the case $\gamma>0$ which corresponds to the g-wave pairing in
the cuprate ground state. Minimum of the free energy (\ref{FSQ}) at fixed supercurrent 
(\ref{jSQ}) is defined by the equations
\begin{eqnarray}
\label{A1A2}
\left. \frac {\partial F_{\Omega }}{\partial A_1}\right|_{j=const}&=&
\left[2A_1K_\gamma\left|\Psi\right|^2+
\lambda_d\cdot
\frac{A_1-A_2\cos(\theta_1-\theta_2)}
{A_1\cos\theta_1-A_2\cos\theta_2}\left|\Psi_s\Psi\right|\right]=0,\\
\left. \frac{\partial F_{\Omega}}{\partial A_2}\right|_{j=const}&=&
\left[2A_2K_\gamma\left|\Psi\right|^2+\lambda_d\cdot
\frac{A_2-A_1\cos(\theta_1-\theta_2)}
{A_1\cos\theta_1-A_2\cos\theta_2}\left|\Psi_s\Psi\right|\right]=0.\nonumber
\end{eqnarray}
There are two solutions of these equations:
\begin{eqnarray}
\label{sym}
symmetric: \ \ \ A_1&=&A_2=
-\frac{\lambda_d}{2K_\gamma}\frac{|\Psi_s|}{|\Psi|}
\frac{\sin\left(\pi\Phi/\Phi_0\right)}{\cos\theta},\\
antisymmetric: \ \ \ A_1&=&-A_2=
\frac{\lambda_d}{2K_\gamma}
\frac{|\Psi_s|}{|\Psi|}\frac{\cos\left(\pi\Phi/\Phi_0\right)}{\sin\theta}.
\nonumber
\end{eqnarray}
The supercurrents and free energies corresponding to symmetric and
antisymmetric solutions are
\begin{eqnarray}
\label{symjf}
j^{(+)}=\frac{2 e \Lambda_d}{\hbar}
\sin^2\left(\pi \frac{\Phi}{\Phi_0}\right)\tan \theta,&&
\hspace{0.1cm}
F_{\Omega}^{(+)}
= \frac{\Lambda_d}{2}\left[-\sin^2\left(\pi\frac{\Phi}{\Phi_0}\right)
+{{2j^2}\over{\left(2e\Lambda_d/\hbar\right)^2
\sin^2\left(\pi \frac{\Phi}{\Phi_0}\right)}}\right],\\
j^{(-)}=-\frac{2e\Lambda_d}{\hbar}
\cos^2\left(\pi \frac{\Phi}{\Phi_0}\right)\mbox{ctan} \theta,&&
\hspace{0.1cm}
F_{\Omega}^{(-)}= \frac{\Lambda_d}{2}\left[-\cos^2\left(\pi\frac{\Phi}{\Phi_0}\right)
+{{2j^2}\over{\left(2e\Lambda_d/\hbar\right)^2
\cos^2\left(\pi \frac{\Phi}{\Phi_0}\right)}}\right],\nonumber
\end{eqnarray}
where $\Lambda_d=\lambda_d^2 |\Psi_s|^2\left/K_\gamma\right.$. The indexes (+), (-)
denote symmetric and antisymmetric solutions correspondingly.
The physical state corresponds to minimum of the free energy and therefore
at $-1/4+m < \Phi/\Phi_0<1/4+m$  the SQUID is in the antisymmetric state
($A_1=-A_2$), and at $1/4+m<\Phi/\Phi_0< 3/4+m$ it is in the symmetric state 
($A_1=A_2$), m is an integer.
One has to remember that eqs. (\ref{sym}) and (\ref{symjf}) are derived at
$A_i\ll 1$, and  they are not valid when $\theta $ is approaching 
$\pi/2$ for the symmetric solution and  when $\theta$ is approaching
$0$ for the antisymmetric solution. 
So at $\gamma > 0$ inside the superconductor
we have g-wave pairing, but in layers of width $l_\gamma$ near the contacts
there is admixture of the d-wave component. If the current is increasing the
admixture is increasing too. So the current pumps the g-wave into the d-wave state 
in the layer of width $l_\gamma$. At the  critical current there is only the d-wave
at the contact. 
The critical currents corresponding to symmetric and 
antisymmetric solutions can be easily found from eq.(\ref{jSQ}) if we
substitute $A_1=A_2=\pi$ or $A_1=-A_2=\pi$.
\begin{eqnarray}
\label{jc+-}
j_c^{(+)}&=&\frac{8e}{\hbar}\left|\Psi\Psi_s\lambda_d
\sin\left(\pi\frac{\Phi}{\Phi_0}\right)\right|,\\
j_c^{(-)}&=&\frac{8e}{\hbar}\left|\Psi\Psi_s\lambda_d
\cos\left(\pi\frac{\Phi}{\Phi_0}\right)\right|.\nonumber
\end{eqnarray}
The real critical current is $j_{cg}=\max\left\{j_c^{(+)}, j_c^{(-)}\right\}$
The plot of $j_{cg}$ as a function of magnetic flux is given at Fig.4b.
It is interesting that it has a period $\Phi_0/2$ which could be naively
interpreted as flux quantization with an effective charge $4e$.

So we see that the dependence of the SQUID critical current on the magnetic flux 
is different for different ground states of the cuprate superconductor.
We considered above the geometry with contact planes parallel to crystal axes
{\bf a} and {\bf b} of the cuprate. If the bulk ground state of the cuprate
has d-wave pairing ($\gamma < 0$) the system remains in the d-wave everywhere 
including the contacts and the critical current is proportional to 
$\left|\sin\left(\pi\Phi/\Phi_0\right)\right|$.
If the bulk ground state of the cuprate has g-wave pairing ($\gamma > 0$) then
the current pumps the g-wave into the d-wave in the layers of the width $l_{\gamma}$
near the contacts. The critical current in this case has a very unusual
dependence on the magnetic flux presented at Fig.4b.\\

Let us consider now a more general geometry of SQUID contacts when
$\lambda_g\neq 0$. We assume that it is a straight angle
rotated with respect to crystal axes of cuprate, therefore
$\lambda_{d1}=-\lambda_{d2}=\lambda_d$, $\lambda_{g1}=\lambda_{g2}=\lambda_g$, 
see Fig.5. We assume also that the angle of rotation is not very small, so 
that $\lambda_d \sim \lambda_g$. Consider at first the case of the d-wave
ground state of the cuprate superconductor ($\gamma < 0$, $\Delta \phi_0=0$).
The free energy is given by the sum of eqs.(\ref{FO}) corresponding to the two 
contacts. The same is valid for supercurrent (\ref{jff}).
Minimization of the free energy at fixed supercurrent with respect to variation
of phases $A_i$ gives in this case a unique solution. At $A_i \ll 1$ the
solution can be found analytically
\begin{equation}
\label{Adl}
A_1=A_2=\frac{\lambda_g}{2K_\gamma}\frac{|\Psi_s|}{|\Psi|}
\frac{\cos\left(\pi\Phi/\Phi_0\right)}{\cos\theta}.
\end{equation}
So at $\lambda_g \ne 0$ external current pumps the d-wave into the g-wave in the
layer of width $\l_{\gamma}$ near the contacts. Keeping in mind that
$A_1=A_2=A$ we can write the supercurrent at arbitrary $A$ in the form
\begin{equation}
\label{jj}
j=j_{c}\cdot \cos (\alpha+A/2)\cdot \sin \theta,
\end{equation}
with $\sin \alpha=\lambda_g\cos(\pi\frac{\Phi}{\Phi_0})\left/
\sqrt{\lambda_d^2\sin^2(\pi\frac{\Phi}{\Phi_0})+
\lambda_g^2\cos^2(\pi\frac{\Phi}{\Phi_0})}\right.$, and the critical current given by
\begin{equation}
\label{jcd}
j_{c}=\frac{8e}{\hbar}|\Psi\Psi_s|\sqrt{\lambda_d^2\sin^2(\pi\frac{\Phi}{\Phi_0})+
\lambda_g^2\cos^2(\pi\frac{\Phi}{\Phi_0})}.
\end{equation}
The plot of the critical current is given at Fig.4c.
The admixture of the g-wave at the contact at critical current is defined by 
$\alpha$. It is equal to 100\% at $\Phi=0$, and it is equal to 0 at $\Phi=\Phi_0/2$.

Now we consider the last situation: geometry with $\lambda_{d1}=-\lambda_{d2}=
\lambda_d$, $\lambda_{g1}=\lambda_{g2}=\lambda_g \ne 0$ (see Fig.5) and
the g-wave ground state of the cuprate superconductor ($\gamma > 0$,
$\Delta \phi_0=\pi$). Minimization of the free energy shows that only one solution
exists, which at small $A_i$ has the form
\begin{equation}
\label{sym1}
A_1=A_2=-\frac{\lambda_d}{2K_\gamma}\frac{|\Psi_s|}{|\Psi|}
\frac{\sin\left(\pi\Phi/\Phi_0\right)}{\cos\theta}.
\end{equation}
It is similar to the symmetric solution (\ref{sym}) for the
geometry with $\lambda_g=0$. 
It is interesting that in this case there is no an 
analog of the antisymmetric solution (\ref{sym}). 
So, as usual the external current pumps the g-wave into the d-wave at the
contact surfaces. The supercurrent at an arbitrary $A_i$, but
with $A_1=A_2=A$ can be written as
\begin{equation}
\label{jjg}
j=j_{c}\cdot \sin (\alpha+A/2)\cdot \sin \theta,
\end{equation}
with exactly the same $j_c$ and $\alpha$ as those for the d-wave ground state,
given by eqs. (\ref{jj}), (\ref{jcd}). The admixture of the d-wave at the contact 
at critical current is defined by $\alpha$. It is equal to 100\% at $\Phi=\Phi_0/2$ 
and it is equal $0$ at $\Phi=0$. So the critical current through the SQUID is
given by eq.(\ref{jcd}) and it is independent of whether we have the d-wave
or the g-wave pairing in ground state. The plot of the critical current is
given at Fig.4c.

The following question arises. In the case of contact planes 
parallel to the crystal axes {\bf a} and {\bf b} the
dependences of critical current on flux are different for different ground 
states: the dependence is given by Fig. 4a for the d-wave ground state, and it is 
given by Fig. 4b for  the g-wave ground state.
On the other hand as soon as the contact is rotated (Fig. 5) the dependence
is the same for the d- and g-wave ground states, and it is given at Fig. 4c.
What is the angle of rotation necessary to change the regime?
The origin of the difference in regimes is in the fact that at $\lambda_g=0$
there are both symmetric and antisymmetric solutions, see eq. (\ref{sym}),
but at finite $\lambda_g$ the anisymmetric solution disappears, 
see eq. (\ref{sym1}). 
One can prove that with increasing $\lambda_g$ from zero the free energy
corresponding to the antisymmetric solution is increasing very fast, and at
$\lambda_g\left/\lambda_d\right. > \beta_c$,
\begin{equation}
\label{bet}
\beta_c \sim \frac{\lambda_d}{2K_\gamma}\frac{|\Psi_s|}{|\Psi|} \ll 1,
\end{equation}
it becomes very high. So the regime is changed when the rotation angle $\beta$
is larger than $\beta_c$. The value of $\beta_c$ is very small because it
is proportional to the tunneling amplitude.

\section{Conclusions}
We have considered the scenario with the small Fermi surface 
consisting of the hole pockets. The picture can be relevant
to underdoped cuprate superconductors. The small Fermi surface together
with mechanism of the magnetic pairing results in the possibility
of having both the d- and the g-wave pairing. Energy splitting
between these states is small. The ground state symmetry is defined
by the interplay between the magnetic pairing and Coulomb repulsion.
We have formulated an effective Ginsburg-Landau theory describing
this situation.

We demonstarate that in the Josephson junction or the SQUID consisting of cuprate
superconductor and conventional superconductor the supercurrent
pumps the d-wave into g-wave (or the g-wave into d-wave) in the thin layer near the
contact. This pumping influences the SQUID interference picture.
If the bulk ground state has  d-wave symmetry and contact planes are
parallel to crystal axes {\bf a} and {\bf b} of the cuprate the dependence
of the SQUID critical current on the magnetic flux is shown at Fig.4a.
This is a well known picture for the d-wave superconductor.
If the bulk ground state has  g-wave symmetry and contact planes are
parallel to crystal axes {\bf a} and {\bf b} of the cuprate the dependence
of the SQUID critical current on the magnetic flux is shown at Fig.4b.
It is very unusual and could be naively interpreted as flux quantization with an 
effective charge $4e$.
If the contact planes are rotated with respect to crystal axes {\bf a} and {\bf b} 
at the angle $\beta > \beta_c$, then the SQUID interference picture is 
the same for the d- and g-wave pairing in the bulk ground state, and this
picture is presented at Fig. 4c.
The angle $\beta_c$ given by eq.({\ref{bet}) is very small, since it is proportional
to the tunneling amplitude.

\section{Acknowledgments}

We are very grateful to D. van der Marel, D. Khomskii, and M. Kuchiev 
for stimulating discussions.
This work was supported by a grant from Australian Research Council.

\newpage

FIIGURE CAPTIONS\\

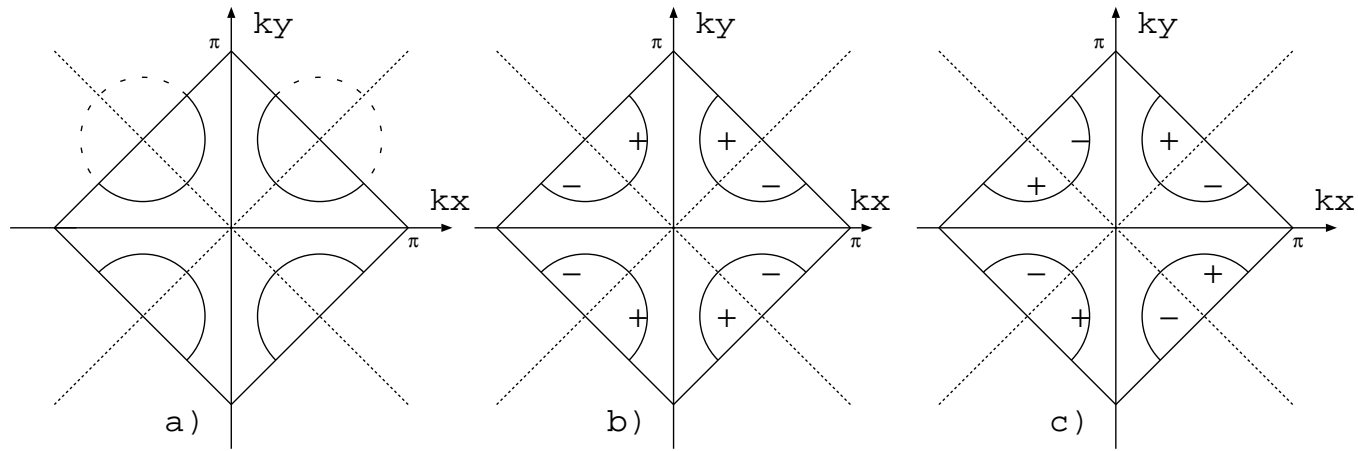
Fig. 1. {\bf a.} Fermi surface in magnetic Brillouin zone which is equivalent to
the two-poket Fermi surface (dashed line).
{\bf b.} Symmetry of the d-wave pairing in momentum space.
{\bf c.} Symmetry of the g-wave pairing in momentum space.\\

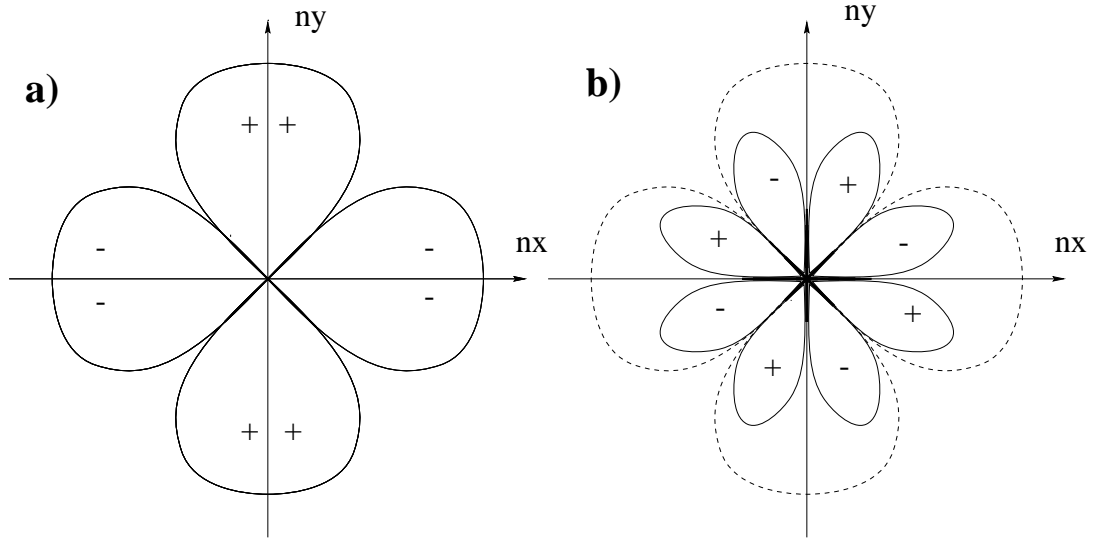
Fig. 2. {\bf a.} The d-wave tunneling amplitude.
{\bf b.} The g-wave tunneling amplitude, $n_x$ and $n_y$ are crystal axes of cuprate.\\

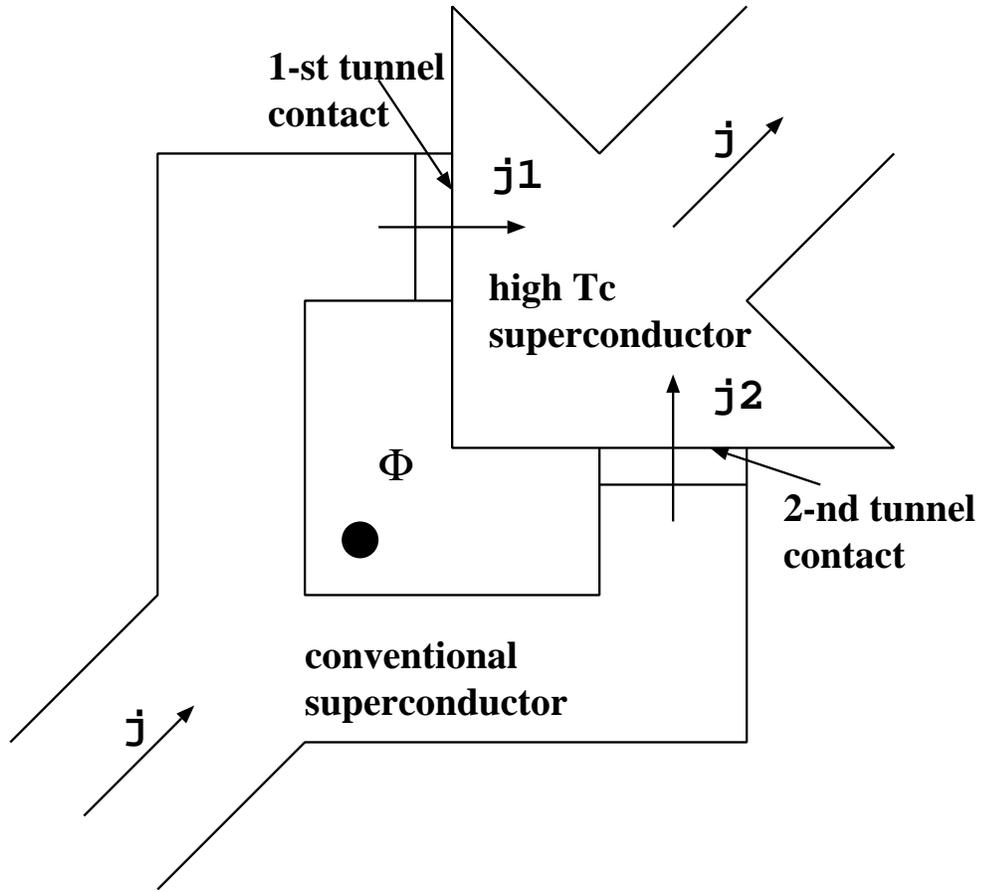
Fig. 3. Geometry of the SQUID experiment with contact planes parallel to the crystal axes.\\

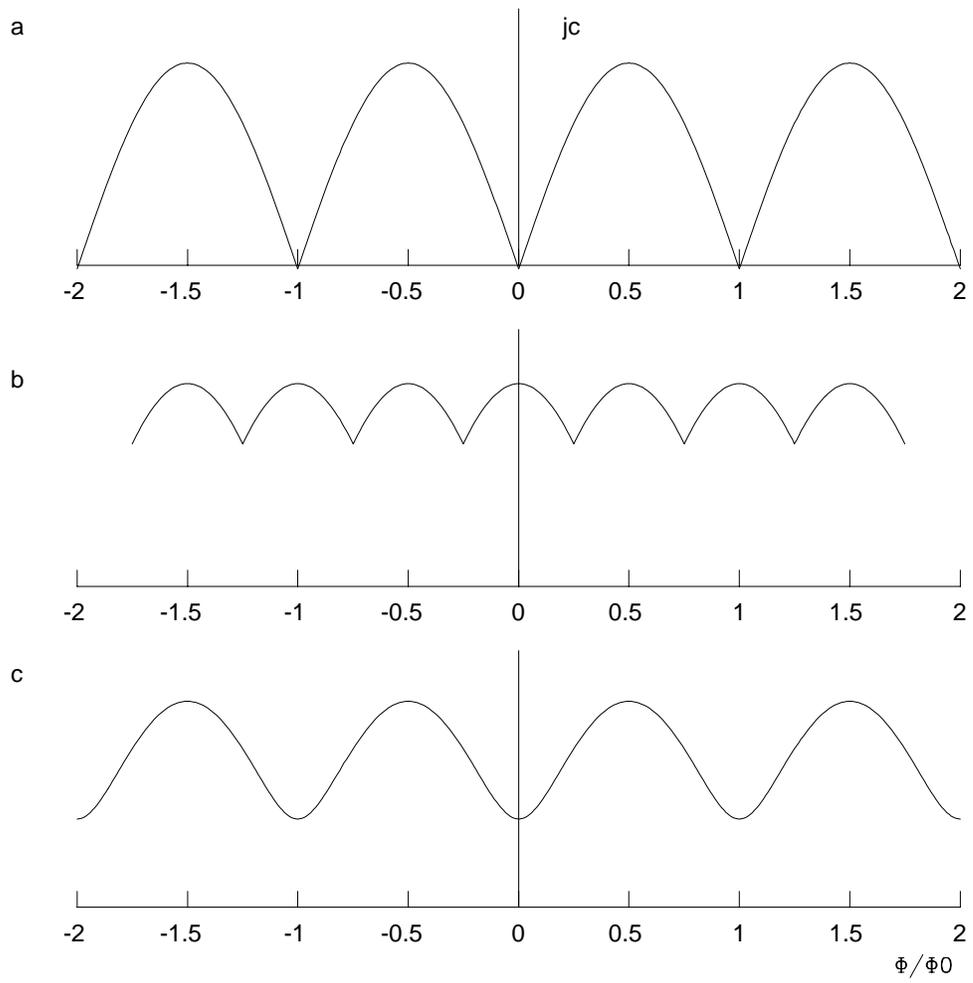
Fig. 4. Dependence of the SQUID critical current on magnetic flux.
{\bf a.} The d-wave bulk ground state and contact planes exactly
parallel to the axes {\bf a} and {\bf b} of the cuprate.
{\bf b.} The g-wave bulk ground state and contact planes exactly
parallel to the axes {\bf a} and {\bf b} of the cuprate.
{\bf c.} The d- or g-wave bulk ground state and superconducting corner rotated
by a angle $\beta>\beta_c$  with respect to the crystal axes {\bf a} and {\bf b}.\\

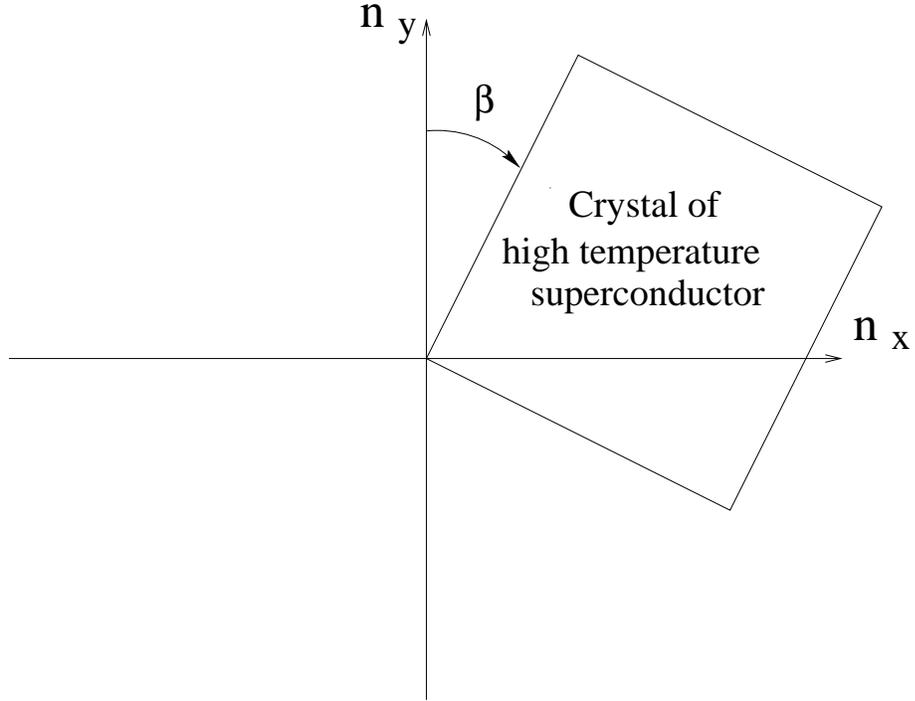
Fig. 5. Geometry of the SQUID experiment with the superconducting corner rotated by an angle $\beta$ with respect to crystal axes.
\end{document}